\documentclass[aps,prb,twocolumn,superscriptaddress,showpacs]{revtex4}
\usepackage{graphicx}
\usepackage{mathrsfs}
\usepackage{bm}
\usepackage{amsmath}
\usepackage{dcolumn}
\usepackage{epstopdf}
\usepackage{dsfont}
\usepackage{amssymb}
\usepackage{tabularx}
\usepackage{array}
\usepackage{colordvi}

\begin{document}
\title{Transmission spectra and valley processing of graphene and carbon nanotube superlattices with inter-valley coupling}
\author{Fuming Xu}
\affiliation{College of Physics and Energy, Shenzhen University, Shenzhen 518060, China.}
\author{Zhizhou Yu}
\affiliation{Department of Physics and Center of Theoretical and Computational Physics, The University of Hong Kong, Hong Kong, China}
\author{Yafei Ren}
\affiliation{ICQD, Hefei National Laboratory for Physical Science at Microscale, and Synergetic Innovation Center of Quantum Information and Quantum Physics, University of Science and Technology of China, Hefei, Anhui 230026, China}
\affiliation{CAS Key Laboratory of Strongly-Coupled Quantum Matter Physics, and Department of Physics, University of Science and Technology of China, Hefei, Anhui 230026, China}
\author{Bin Wang}
\affiliation{College of Physics and Energy, Shenzhen University, Shenzhen 518060, China.}
\author{Yadong Wei}
\email[Correspondence author:~]{ywei@szu.edu.cn}
\affiliation{College of Physics and Energy, Shenzhen University, Shenzhen 518060, China.}
\author{Zhenhua Qiao}
\email[Correspondence author:~]{qiao@ustc.edu.cn}
\affiliation{ICQD, Hefei National Laboratory for Physical Science at Microscale, and Synergetic Innovation Center of Quantum Information and Quantum Physics, University of Science and Technology of China, Hefei, Anhui 230026, China}
\affiliation{CAS Key Laboratory of Strongly-Coupled Quantum Matter Physics, and Department of Physics, University of Science and Technology of China, Hefei, Anhui 230026, China}

\begin{abstract}
  We numerically investigate the electronic transport properties of graphene nanoribbons and carbon nanotubes with inter-valley coupling, e.g., in $\sqrt{3}N \times \sqrt{3}N$ and $3N \times 3N$ superlattices. By taking the $\sqrt{3} \times \sqrt{3}$ graphene superlattice as an example, we show that tailoring the bulk graphene superlattice results in rich structural configurations of nanoribbons and nanotubes. After studying the electronic characteristics of the corresponding armchair and zigzag nanoribbon geometries, we find that the linear bands of carbon nanotubes can lead to the Klein tunnelling-like phenomenon, i.e., electrons propagate along tubes without backscattering even in the presence of barrier. Due to the coupling between $K$ and $K^\prime$ valleys of pristine graphene by $\sqrt{3} \times \sqrt{3}$ supercell, we propose a valley-field-effect transistor based on the armchair carbon nanotube, where the valley polarization of the current can be tuned by applying a gate voltage or varying the length of the armchair carbon nanotubes.
\end{abstract}

\maketitle

\section{introduction}
Valleytronics aim to design high-efficiency and low-dissipation electronic devices by manipulating the Bloch electrons' valley degree of freedom, which refers to the local minima of the electronic band structure in the reciprocal space. In some traditional multi-valley systems, such as silicon,\cite{silicon1,silicon2,silicon3} bismuth,\cite{bismuth} and diamonds,\cite{diamond} the valley degree of freedom is shown to be controllable to carry and transport information. In two-dimensional materials, honeycomb-lattice systems are of special interest in the study of valleytronics due to the presence of two inequivalent valleys $K$ and $K'$.\cite{graphene3,graphene4,graphene5} Particularly, \textit{graphene} has attracted much attention due to its excellent electronic and mechanical properties \cite{graphene1,graphene11,graphene2} Various valleytronics devices have been proposed in graphene nanostructures\cite{valley061,valley062,valley071,valley072,valley11} utilizing, e.g., zigzag edges,\cite{valleygraphene1} zero-line modes,\cite{ZLM-1,ZLM-partition,review} topological line defects,\cite{valleygraphene21,valleygraphene22,valleygraphene23} strain and mechanical engineering,\cite{valleygraphene31,valleygraphene32} as well as temperature gradient,\cite{valleygraphene4} to generate and control valley-polarized currents.

Recently, a new valley engineering mechanism is proposed in $\sqrt{3}N \times \sqrt{3}N$ or $3N \times 3N$ superlattices of graphene.\cite{qiao15} Due to the band folding in the superlattice, the inequivalent $K/K'$ valleys in pristine graphene are folded into the same $\Gamma$ point and thus inter-valley coupling arises that act as valley-orbit coupling similar to spin-orbit coupling providing promising valley-processing mechanisms via electrical means. Reference~[\onlinecite{qiao15}] suggested that the $\sqrt{3}N \times \sqrt{3}N$ or $3N \times 3N$ superlattices could be realized in periodically doped graphene. More recently, such kind of special supercells are also shown to appear in graphene proximity-coupled with topological insulator substrates.\cite{arxivTI}

In this article, we explore the potential application of these graphene superlattices in valleytronics and extend our study to carbon nanotubes. Within top-adsorption case, we study the representative $\sqrt{3} \times \sqrt{3}$ superlattice of graphene and carbon nanotubes without loss of generality since the inter-valley coupling mechanisms are universal features in these superlattices. The $\sqrt{3} \times \sqrt{3}$ superlattice with top-adsorption introduces multiple structural configurations of graphene nanoribbons and nanotubes. We focus on three kinds of zigzag ribbons, two kinds of armchair ribbon, and typical armchair and zigzag single walled nanotubes, and theoretically investigate their electronic properties using the tight-binding model. Our numerical results show that there exist Klein tunneling-like phenomena in $\sqrt{3} \times \sqrt{3}$ armchair and zigzag carbon nanotubes even in the presence of barrier. By employing the inter-valley coupling to induce valley processing, we propose a valley-field-effect transistor consisting of pristine and $\sqrt{3}\times \sqrt{3}$ armchair carbon nanotube, which generates nearly fully valley-polarized current at large gate voltage and electron energy scale.

The remaining of our paper is organized as follows. In Sec.~\ref{sec2}, the tight-binding Hamiltonian of the bulk $\sqrt{3}\times \sqrt{3}$ graphene and the Green's function method are introduced. The numerical results on electronic and transport properties of various confined graphene nanostructures, including zigzag and armchair ribbons, and single walled nanotubes are shown in Sec.~\ref{sec3}. A brief summary is given in Sec.~\ref{sec4}.

\section{Tight-binding Hamiltonian and theoretical formalism}\label{sec2}
In $\sqrt{3} \times \sqrt{3}$ graphene superlattice, carbon atoms are still the majority and $\pi$-orbital expansion is therefore a reasonable starting point. The nearest-neighbor tight-binding Hamiltonian of the bulk $\sqrt{3}\times \sqrt{3}$ graphene reads:\cite{qiao15}
\begin{eqnarray}
H = -\sum_{\langle i,j \rangle}  t_{i,j} \,\,  c_i^{\dag} c_j
+ \sum_{i} u_{i} \, \, c_{i}^{\dag} c_{i}, \label{eq01}
\end{eqnarray}
where $c^{\dag}_i$($c_{i}$) is the $\pi$-orbital creation (annihilation) operator on site $i$. To recover the single-valley metallic phase of bulk $\sqrt{3} \times \sqrt{3}$ graphene superlattice,\cite{qiao15} the system parameters are precisely determined. The nearest-neighbor hopping amplitudes $t_{i,j}$ are $t_{1}=2.9$~eV between top-absorption site and adjacent carbon atoms, and $t_2=2.6$~eV between carbon atoms, respectively. The on-site potentials $u_{i}$ of different sites are chosen to be $u_1=-4.79$~eV for top-absorption site, $u_2=-1.35$~eV for its three nearest carbon sites, and $u_3=-1.05$~eV for the rest carbon sites with no top-absorption neighbors. These parameters perfectly recover the band structure in Ref. [\onlinecite{qiao15}], where periodic adatom or top-absorption introduces symmetry-breaking and valley-scattering in graphene superlattice. Detailed procedures on parameter selection can be found in Ref. [\onlinecite{qiao15}].

\begin{figure}[tbp]
\centering
\includegraphics[width=\columnwidth]{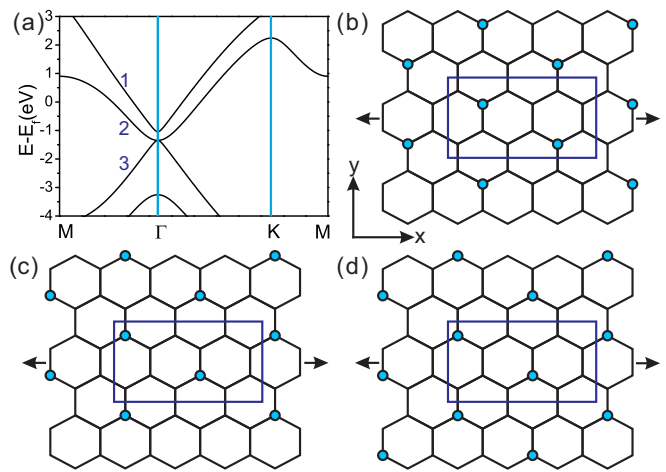}
\caption{Panel (a): band structure of bulk $\sqrt{3} \times \sqrt{3}$ graphene superlattice calculated from tight-binding Hamiltonian defined in Eq.\ref{eq01} with periodic boundary conditions. Panel (b), (c) and (d): schematic plots of three types of $\sqrt{3} \times \sqrt{3}$ zigzag graphene ribbons. Green dots stand for top-absorption sites in the lattice. Blue rectangles indicate the unit cell in every setup. The ribbons are finite in $y$ direction and periodically infinite in $x$ direction. }
\label{fig01}
\end{figure}

Based on this tight-binding Hamiltonian, the band structure of bulk $\sqrt{3} \times \sqrt{3}$ graphene is obtained as shown in Fig.~\ref{fig01}(a) where the inequivalent $K$ and $K'$ valleys in pristine graphene are folded at the same $\Gamma$ point due to the band folding. In our transport study, we focus on an energy interval within several eV around the Fermi energy. In Fig.\ref{fig01}(a), it is obvious that the three bands (labeled as "1", "2" and "3") dominate in this energy range. Bands $2$ and $3$ form an ideal quadratic crossover.\cite{qiao15} Moreover, one can also notice that bands "1" and "3" are almost linear around $\Gamma$ point. The influence of such kind of linear dispersion will be discussed when studying the transport properties of $\sqrt{3}$ graphene nanoribbons and carbon nanotubes.

In below, we utilize the Green's function method to investigate the valley-related electronic transport properties. From the tight-binding Hamiltonian shown in Eq.~(\ref{eq01}), the transmission coefficient of electrons at energy $E$ can be expressed as:
\begin{equation}
T(E) = {\rm Tr}[\Gamma_L G^r \Gamma_R G^a],
\label{eq02}
\end{equation}
where $G^r=[E-H-\Sigma^r]^{-1}$ is the retarded Green's function, and $G^a=(G^r)^\dagger$ is the advanced Green's function. $\Sigma^r = \Sigma^r_L + \Sigma^r_R$ is the self-energy from the left and right leads. $\Gamma_{L/R}$ is the line width function describing the coupling between the left/right lead and the central scattering region, and can be defined as $\Gamma_{L/R}=i[\Sigma^{r}_{L/R} -\Sigma^{a}_{L/R}]$.

\section{Numerical results and discussions}\label{sec3}

In this Section, numerical results on the electronic structures and transport properties of typical nanoribbons and nanotubes of $\sqrt{3} \times \sqrt{3}$ graphene supercell are presented, including both zigzag and armchair geometries.

\subsection{Zigzag ribbons of $\sqrt{3} \times \sqrt{3}$ graphene superlattice}\label{sec31}

As displayed in Fig.~\ref{fig01}, we exhibit three kinds of $\sqrt{3}\times \sqrt{3}$ zigzag graphene nanoribbons (ZGRs). Top-absorption sites are highlighted as green dots. Blue rectangles are used to indicate the unit cells in different configurations. For simplicity, the setups in panels (b), (c), and (d) of Fig.~\ref{fig01} are respectively denoted as ZGR-1, ZGR-2, and ZGR-3. Their geometric differences can be easily distinguished from the relative position of top-absorption sites and their presence/absence at the ribbon boundaries. The lower boundaries of ZGR-1 and ZGR-2 are purely consisted of carbon atoms while the absorption sites of these two samples are locate on different sublattices. Differently, adsorption sites appear at the lower boundary of ZGR-3. As displayed in Fig.~\ref{fig01}, these ZGRs have finite widths along $y$ direction and are periodic along $x$ direction. These zigzag ribbons can be realized by cutting a large $\sqrt{3}\times \sqrt{3}$ graphene sheet along proper directions, like the fabrication of graphene nanoribbon\cite{cutgraphene1} or through the lithography method.\cite{cutgraphene2} In our first principles calculation,\cite{yu14} we found that ZGR-1 exhibits the lowest Gibbs free energy, while ZGR-3 exhibits the highest one. Therefore, ZGR-1 should theoretically be the most stable structure in these ZGRs. In our following discussions, all these setups have been considered and we will show that they possess distinct electronic properties.

\begin{figure}[tbp]
\centering
\includegraphics[width=\columnwidth]{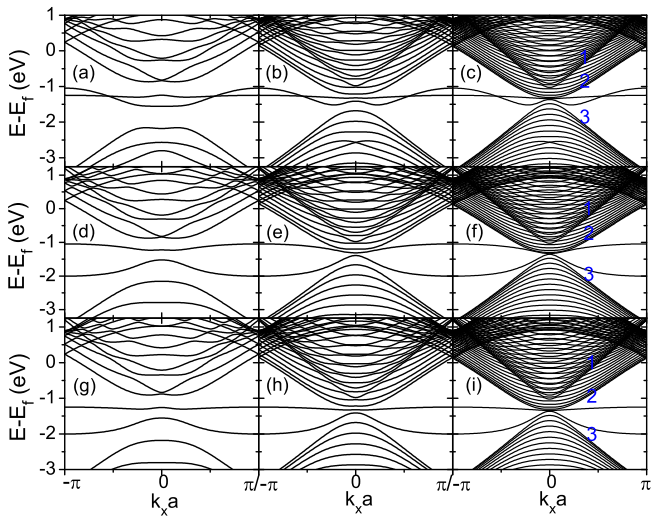}
\caption{Band structures of three types of $\sqrt{3}$ zigzag ribbons with various system sizes. Panels (a)-(c): band structures of ZGR-1 with ribbon width 24, 52, and 100 lattice sites; Panels (d)-(f): dispersion relations of ZGR-2 with system size 24, 52, and 100 sites; Panels (g)-(i): bands of ZGR-3 with width 24, 52, and 100 sites.}
\label{fig02}
\end{figure}

Figure~\ref{fig02} displays the band structures of three different zigzag graphene ribbons for different widths (e.g., 24, 52, and 100 lattice sites). The first, second and third rows of Fig.~\ref{fig02} correspond respectively to ZGR-1, ZGR-2 and ZGR-3. One can find that Fig.~\ref{fig02}(a) (for ZGR-1 with a width of 24 lattice sites) exhibits a large energy gap $\Delta$ located in the interval of [$-1$, $-2$]~eV, deeply underneath the Fermi energy. Meanwhile, two nearly flat bands lie in the gap. By projecting the local density of states of the two bands onto the lattice sites, we find that they are localized along the zigzag ribbon boundaries and thus they are edge modes, which originate from the dangling bonds along the zigzag edges, like those in zigzag ribbon of pristine graphene. As clearly presented in Fig.~\ref{fig01}, the upper and lower edges of these ZGRs are different, leading to the formation of two different edge modes. As the ribbon width increases, the energy gap above the edge modes vanishes first as shown in Fig.~\ref{fig02}(b) and the lower gap closes at a larger ribbon width as displayed in Fig.~\ref{fig02}(c). Therefore, these two band gaps arise from the finite-size effect and disappear when the system size is sufficiently large.

For ZGR-1 as the system width increases, another observation is that the edge modes are pinned at $k_x = \pm \pi$ but evolve with the bulk states around $k_x \approx 0$. One can also notice that, in Fig.~\ref{fig02}(c), the band structure establishes three envelopes, labeled as ``1"-``3". Upon larger system size, the envelopes become further enhanced with higher energy degeneracies. Compared with Fig.~\ref{fig01}(a), one reasonable explanation is that these envelopes reflect the three bands in the corresponding bulk bands. For ZGR-2, bands for system widths 24, 52, and 100 sites are respectively plotted in Fig.~\ref{fig02}(d), (e) and (f), which exhibit similar characters as those in ZGR-1 except that these two edge modes are separable and divide the band gap into three narrow ones. As the ribbon width increases, the upper and lower gaps close first and the middle gap disappears at last. The cases for ZGR-3 is similar to that of ZGR-2 as displayed in the last three panels of Fig.~\ref{fig02}. However, different from ZGR-2, the upper surface band of ZGR-3 is more flat. Nevertheless, the band structures of these three types of zigzag ribbons share some common features, including surface bands pinning at $k_x = \pm \pi$ and bands envelopes at large system sizes. Besides, these zigzag ribbons are all metallic since the Fermi level lies in the conduction band.

\begin{figure}[tbp]
\centering
\includegraphics[width=\columnwidth]{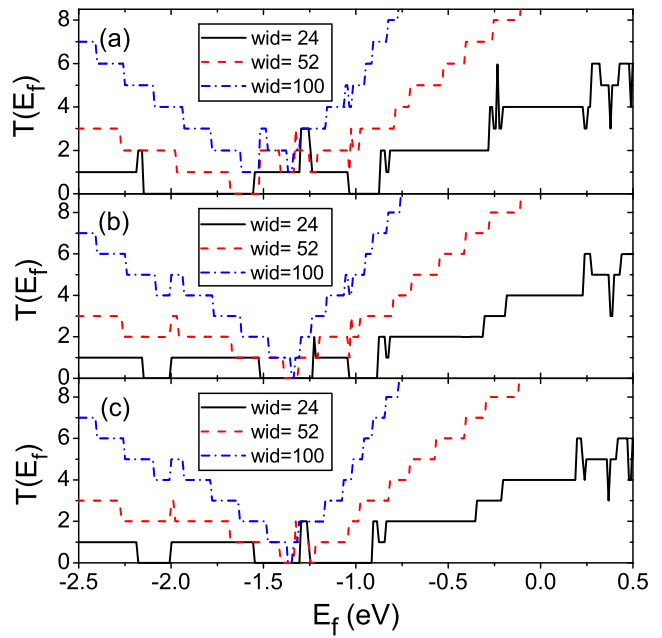}
\caption{Transmission coefficient as a function of Fermi energy for three types of $\sqrt{3}\times\sqrt{3}$ zigzag graphene ribbons (ZGRs) at different widths. Panel (a), (b), and (c) corresponds to ZGR-1, ZGR-2, and ZGR-3, respectively. In all panels, three ribbon widths 24, 52, and 100 sites are evaluated. }
\label{fig03}
\end{figure}

In Fig.~\ref{fig03}, we plot the transmission coefficients of these zigzag ribbons as a function of energy, where the left and right leads are exactly extended from the central region, hence resulting in the quantized $T(E_{\rm F})$. Panels (a), (b) and (c) correspond respectively to ZGR-1, ZGR-2 and ZGR-3. In each setup, three ribbon widths (i.e., 24, 52 and 100 lattice sites) are considered. One can see an exact mapping between the transmission coefficients in Fig.~\ref{fig03} and the band structures in Fig.~\ref{fig02}. At small system size, all the ZGRs have zero transmission coefficients at certain energy regions below the Fermi level, where the dispersion relations show energy gaps. For ZGR-1, there are two zero-transmission-coefficient regions while three gaps exist in ZGR-2 and ZGR-3 at ribbon width (i.e., with 24 lattice sites in black lines). At system size of 52, the zero conducting ranges shrink in all panels of Fig.~\ref{fig03}. Only one gap are present in the ZGR-2 and ZGR-3, and two gaps still reside in the transmission spectrum of ZGR-3, shown in red lines. For a larger system with 100 lattice sites, ZGR-1 has no zero transmission area in the whole energy interval and single narrow gaps appear in panels (b) and (c) of Fig.~\ref{fig03}. These finite-size gaps eventually disappear at even larger systems. As the ribbon width increases, transmission coefficient at the same energy increases rapidly for all systems. We also observe some oscillation behavior of $T(E_{\rm F})$ around $E_{\rm F} \approx$ 0.5~eV, which can be attributed to the band overlapping at these energies. One can deduce from Fig~.\ref{fig02} that, the oscillations tend to be more wild at larger system sizes, which is confirmed by our transport calculations.

\begin{figure}[tbp]
\centering
\includegraphics[width=\columnwidth]{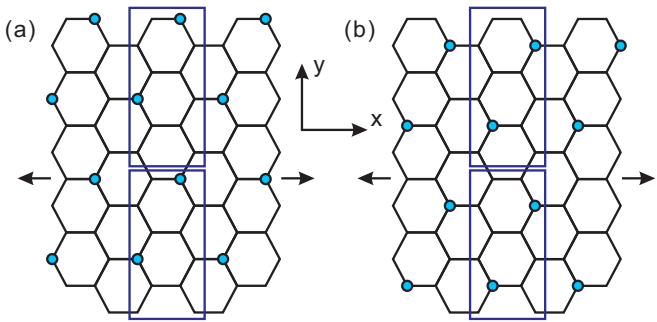}
\caption{Sketches of two types of $\sqrt{3}$ armchair graphene ribbons: AGR-1 and AGR-2, respectively. Blue rectangles show the corresponding unit cells and green dots indicate top-absorption sites. The ribbons have finite widths in $y$ direction and extend to infinite $\pm x$. }
\label{fig04}
\end{figure}

\subsection{Armchair ribbons of $\sqrt{3} \times \sqrt{3}$ graphene superlattice}\label{sec32}
Depending on the relative positions of top-absorption on the honeycomb lattice, there are two different configurations of armchair graphene nanoribbons (AGRs). We schematically plot these AGRs with widths of two full unit cells in Fig.~\ref{fig04}. For simplicity, we denote the armchair ribbon shown in Fig.~\ref{fig04}(a) as AGR-1 and the other one as AGR-2. Unlike zigzag ribbons of $\sqrt{3}\times \sqrt{3}$ graphene, both armchair ribbons reproduce perfect $\sqrt{3} \times \sqrt{3}$ periodicity. The first-principles calculation suggests that the Gibbs free energy of AGR-1 is higher than that of AGR-2.\cite{yu14} Combining with the results of $\sqrt{3}\times \sqrt{3}$ ZGRs, one can conclude that the ribbons for both zigzag and armchair forms are less stable when the top-absorption sites reside on the ribbon boundaries.

\begin{figure}[tbp]
\centering
\includegraphics[width=\columnwidth]{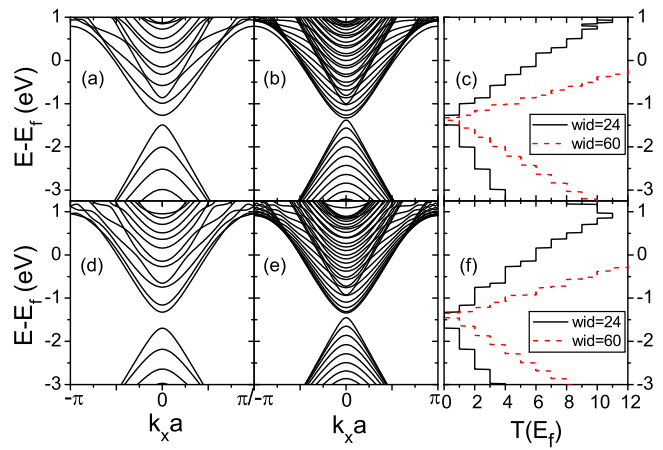}
\caption{Panels (a) and (b): band structures of AGR-1 with system widths 24 and 60 sites. Panel (c) plots the corresponding $T$ vs $ E_{\rm F}$ curves for these system sizes. Panels (d) and (e): band structures of AGR-2 at ribbon widths 24 and 60 sites. Panel (f) shows the transmission of AGR-2.}
\label{fig05}
\end{figure}

The electronic properties of these AGRs are numerically investigated and displayed in Fig.~\ref{fig05}. In our calculations, two ribbon widths (24 and 60 lattice sites) are considered for both AGR-1 and AGR-2. One notices that both armchair ribbons are good conductors, similar as the ZGRs. Figures~\ref{fig05}(a) and \ref{fig05}(b) plot the band structures of AGR-1. Apparently, there is a direct band gap at $k_x = 0$ around $E \sim -1.4$~eV. As the system size increases from 24 to 60 lattice sites, this gap becomes narrower as shown in Fig.~\ref{fig05}(b), indicating its finite-size nature. The transmission coefficient of AGR-1 as a function of energy is displayed in Fig.~\ref{fig05}(c). It is found that the conducting-forbidden region matches exactly the band gaps in the left panels. This gap is gradually reduced when system size increases and finally vanishes as the ribbon width is large enough. For AGR-2, the numerical results are drawn in the lower panels of Fig.~\ref{fig05}. The energy dispersion of AGR-2 is rather similar to that of AGR-1 except that AGR-2 has a larger band gap at the same system size. The transmission coefficient versus energy curves in Fig.~\ref{fig05}(f) also confirm this observation. As a result, a wider ribbon width is required to close the band gap in AGR-2.

Compared with ZGRs, there is no edge mode in AGRs. This is a much natural expectation for armchair-edged ribbons because of the valley mixture behaviour. Another striking difference from ZGRs is that there is less oscillation feature in the transmission spectrum of AGRs, which can be attributed to the absence of band-folding from $1\times1$ to $\sqrt{3}\times \sqrt{3}$ supercells in the armchair configurations. From Figs.~\ref{fig05}(b) and \ref{fig05}(e), one can observe that three band envelopes develop to reproduce the bulk bands of $\sqrt{3} \times \sqrt{3}$ graphene superlattice.

\subsection{Typical single wall carbon nanotubes of $\sqrt{3} \times \sqrt{3}$ graphene supercell}\label{sec33}

The carbon nanotubes of $\sqrt{3} \times \sqrt{3}$ graphene supercell have similar structures as pristine carbon nanotubes. Here we consider two representing configurations: single wall armchair and zigzag carbon nanotubes. Theoretically, $\sqrt{3} \times \sqrt{3}$ armchair carbon nanotubes can be formed by connecting the upper and lower boundaries of ZGR-3 as displayed in Fig.~\ref{fig01}(d) with proper bonding. Meanwhile, $\sqrt{3} \times \sqrt{3}$ zigzag nanotubes can be formed by rolling up armchair ribbons as displayed in Fig.~\ref{fig04} and linking their edges accordingly. Carbon nanotubes constructed in these ways possess the full $\sqrt{3} \times \sqrt{3}$ periodicity. In our consideration, we adopt the classification rule of pristine single wall carbon nanotubes\cite{CNTreview1,CNTreview2} to denote these two types of $\sqrt{3}\times \sqrt{3}$ carbon nanotubes, where notations $(n,n)$ and $(n,0)$ stand for armchair and zigzag tubes, respectively. Integer $n$ in these notations refers to the number of unit vectors defined in the honeycomb lattice of pristine bulk graphene.

\begin{figure}[tbp]
\centering
\includegraphics[width=\columnwidth]{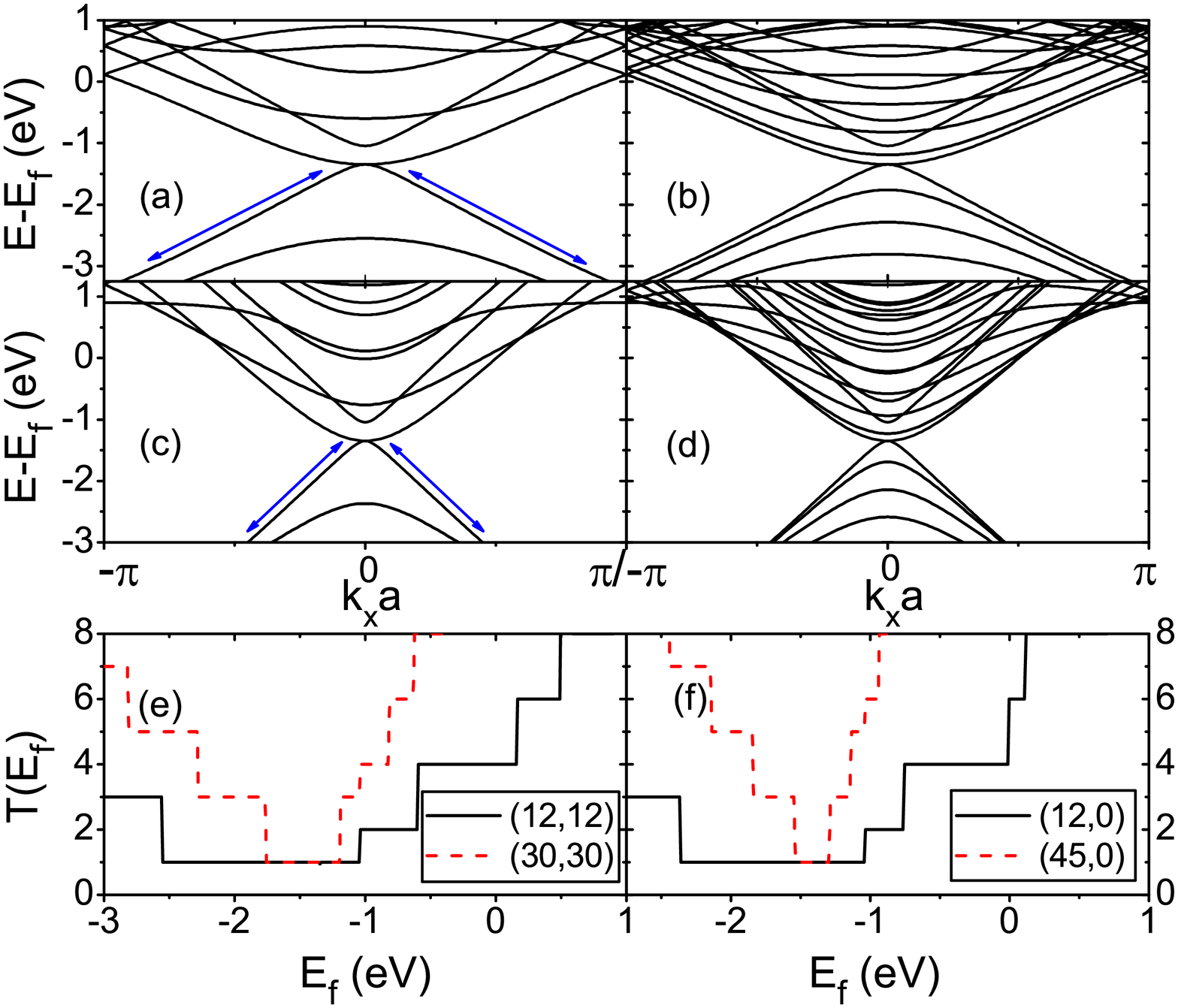}
\caption{Panels (a) and (b): band structures of $\sqrt{3}\times \sqrt{3}$ armchair carbon nanotubes fabricated from ZGR-3 shown in Fig.~\ref{fig01}(d), at tube sizes $(12,12)$ and $(20,20)$. Panels (c) and (d): dispersion relations of $(12,0)$ and $(45,0)$ zigzag carbon nanotubes, formed by folding armchair ribbons (Fig.~\ref{fig04}) in $y$ direction. Panel (e) plots the $T(E_{\rm F})$ curves of $(12,12)$ and $(30,30)$ armchair tubes at different Fermi energies. Panel (f) shows the transmission of zigzag tubes with system sizes $(12,0)$ and $(45,0)$.}
\label{fig06}
\end{figure}

Figure~\ref{fig06} displays the band structures of these $\sqrt{3}\times \sqrt{3}$ armchair and zigzag carbon nanotubes at different tube sizes where one can find that, similar to $\sqrt{3} \times \sqrt{3}$ graphene nanoribbons, these two types of carbon nanotubes are also metallic and their Fermi energies lie deeply into the conduction bands. Specifically, Figs.~\ref{fig06}(a) and \ref{fig06}(b) display the energy dispersion relations of the $(12,12)$ and $(20,20)$ armchair carbon nanotubes. When transforming the zigzag ribbons into armchairs tubes, it is clear that there are neither edge modes nor gaps in the bands of $\sqrt{3}\times \sqrt{3}$ armchair nanotubes. In Fig.~\ref{fig06}(a), one can see that there is a band touching at $E \approx -1.4$~eV. Near the touching point, the band density is low for this $(12,12)$ armchair tube. For a larger $(20,20)$ tube as shown in Fig.~\ref{fig06}(b), the band density becomes denser with a fixed band touching point. The band structures of the two $\sqrt{3}\times \sqrt{3}$ zigzag carbon nanotubes are plotted in Fig.~\ref{fig06}(c) and \ref{fig06}(d). The band touching also appear in these systems. For the same energy, the bands of zigzag tubes under the touching point reside around $\rm k_x = 0$ point, instead of spreading in the whole Brillouin zone as in the armchair tubes. This behavior indicates that current carrier in zigzag ribbons has a larger group velocity in this energy range. Except this minor difference, the bands between armchair and zigzag tubes share lots of similarities.

Figures~\ref{fig06}(e) and \ref{fig06}(f) display the transmission coefficients as a function of energy for the $\sqrt{3}\times \sqrt{3}$ armchair and zigzag carbon nanotubes. The results for $(12,12)$ and $(30,30)$ armchair nanotubes are shown in Fig.~\ref{fig06}(e). In a wide energy range of $E_{\rm F} \in [-2.5,-1)$~eV, there is only single conducting channel in the$(12,12)$ armchair tube. This energy range corresponds to the two touched bands in Fig.~\ref{fig06}(a) for the same system parameters. The increase of system size narrows this $T=1$ region, as one can see from the red line for the $(30,30)$ armchair nanotube. Another interesting finding in Fig.~\ref{fig06}(e) is that the transmission coefficients increase by 2 at lot of energy points. This behaviour reveals that many energy levels in the armchair tubes are doubly degenerate, regardless of the system size. The transmission spectra of two $\sqrt{3}\times \sqrt{3}$ zigzag tubes with sizes of $(12,0)$ and $(45,0)$ are displayed in Fig.~\ref{fig06}(f). Despite their distinct geometric configurations, the $\sqrt{3}\times \sqrt{3}$ zigzag and armchair carbon nanotubes have similar $T(E_{\rm F})$ profiles. We also observe the large single mode region for small tube size and lots of doubly degenerate bands at zigzag nanotubes.

\begin{figure}[tbp]
\centering
\includegraphics[width=0.9\columnwidth]{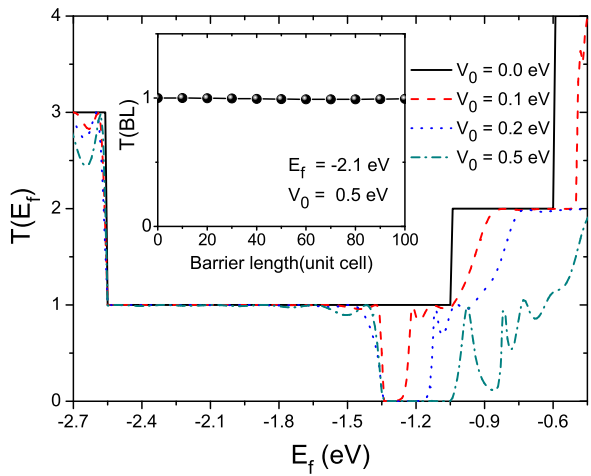}
\caption{Transmission vs Fermi energy of a $\sqrt{3}$ $(12,12)$ armchair carbon nanotube where a potential barrier with a length of 10 unit cells exists. Various colored lines stand for different barrier heights $V_0$. Inset: transmission of the same system at fixed energy $E_{\rm F}=-2.1 eV$ and barrier height $V_0=0.5 eV$ for several barrier lengths. The length scale 'unit cell' is indicated in Fig.\ref{fig01}(d).}
\label{fig07}
\end{figure}

In Figs.~\ref{fig06}(a) and \ref{fig06}(c), the linear bands parallel to the blue arrows are rather attractive. In single layer graphene, the massless Dirac fermion with linear energy dispersion leads to the counterintuitive Klein paradox,\cite{kleinparadox1,kleinparadox2} where incoming relativistic particles penetrate high barrier with nearly perfect transmission at certain angles. Klein tunnelling behaviour has been revealed in various graphene-based systems, such as graphene $p-n$ junctions,\cite{klein1,klein2} deformed single layer,\cite{klein3} and twisted bilayer graphene,\cite{klein4} as well as spin-related graphene system.\cite{klein5} Similar perfect transmission is also observed when electron propagates in pristine carbon nanotubes.\cite{CNTklein1,CNTklein2,CNTklein3} Inspired by these findings, we carry out numerical calculation to show the existence of Klein paradox in the $\sqrt{3} \times \sqrt{3}$ carbon nanotubes.

Our system in consideration is a $\sqrt{3} \times \sqrt{3}$ $(12,12)$ armchair nanotube. A potential barrier with height of $V_0$ and certain length is exerted on the tube while other parts of the system remains unchanged. We calculate the electron transmission coefficient through the barrier. Figure~\ref{fig07} plots the transmission coefficient as a function of energy for several different barrier heights. The barrier length is fixed at 10 unit cells, and this length scale of $\sqrt{3}\times \sqrt{3}$ armchair tube is depicted in Fig.~\ref{fig01}(d), containing 4 sites along the $x$ direction. It is found that, in a wide energy window of $E \in [-2.5,-1.7]$~eV, electrons in these armchair carbon nanotubes can almost perfectly penetrate the potential barrier. Quantized transmission coefficients of $T(E_{\rm F})=1$ can be achieved for barrier height up to $V_0=0.5$~eV. Outside this region, the transmission coefficient dramatically drops in the presence of potential barrier.

We carefully examined the numerical results and found that, the $T(E_{\rm F})=1$ energy window $E \in [-2.5,-1.7]$~eV corresponds to the linear dispersion region in Fig.~\ref{fig06}(a), suggesting a direct correlation between linear dispersion and nearly perfect transmission. In energy range $E \in [-2.5,-1.7]$~eV, the linear dispersion relation guarantees the electrons high group velocity, or so-called relativistic electron. In our carbon nanotube system with potential barrier, the incident electrons normally collide with the barrier. The high-velocity relativistic electron can penetrate the barrier without back-scattering, resulting almost perfect transmission or reflectionless tunneling. We further checked the dependence of this almost perfect tunnelling on barrier length, at fixed electron energy of $E_{\rm F}=-2.1$~eV and barrier strength of $V_0=0.5$~eV. The transmission function versus barrier length is plotted in the inset of Fig.~\ref{fig07}. Regardless of the barrier length, the system with longer barrier remains transparent to the incident electrons and also gives rise to quantized transmission coefficient $T = 1$. These numerical evidences strongly suggest the existence of Klein tunnelling-like behavior in the $\sqrt{3}\times\sqrt{3}$ armchair carbon nanotubes, where electrons transport without backscattering in the system, independent of barrier length and heights up to half an electron volt. Similar perfect tunnelling is also observed in the $\sqrt{3}\times\sqrt{3}$ zigzag carbon nanotubes. However, when the tube size increases, bulk states arise in the whole energy range and gradually co-exist with the linear bands. As a result, the Klein tunnelling-like phenomenon becomes obscure and eventually disappears.

\subsection{Valley processing in $\sqrt{3}\times\sqrt{3}$ armchair nanotubes}\label{sec34}

\begin{figure}[tbp]
\centering
\includegraphics[width=\columnwidth]{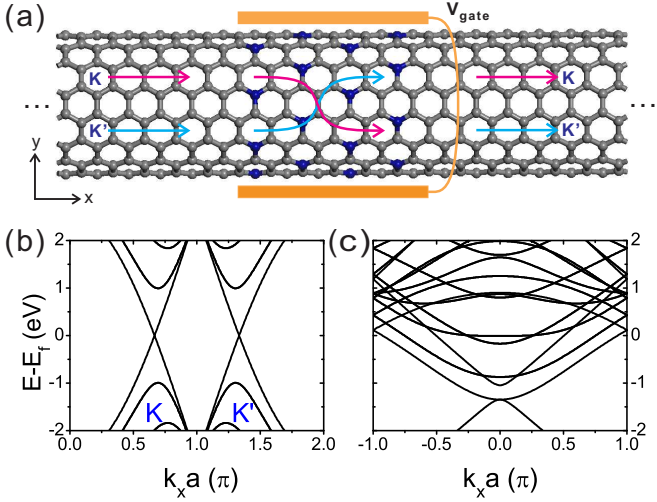}
\caption{Panel (a): Illustration of a valley-field-effect-transistor based on pristine and $\sqrt{3}$ armchair carbon nanotubes. A gate voltage is applied on the $\sqrt{3}$ armchair carbon nanotube region, which functions as valley-processing unit. The $\sqrt{3}$ armchair carbon nanotube is characterized by the blue-colored top-absorption sites. Panels (b) and (c): band structures of $(8,8)$ pristine and $\sqrt{3}\times\sqrt{3}$ armchair carbon nanotubes, respectively.}
\label{fig08}
\end{figure}

Due to the band folding of the $\sqrt{3} \times \sqrt{3}$ superlattice, the inequivalent $K$ and $K'$ valleys in pristine graphene and carbon nanotubes are folded into $\Gamma$ point. Therefore, the intervalley coupling and valley-orbit coupling effects emerge in these nanotubes of $\sqrt{3}\times \sqrt{3}$ superlattices of graphene,\cite{qiao15} which qualify them as potential valley-processing materials. Here, we propose a heterostructure composed of $1\times1$ and $\sqrt{3}\times\sqrt{3}$ armchair carbon nanotubes to act as a valley filter or valley polarizer as illustrated in Fig.~\ref{fig08}(a), which consists of two identical leads made of pristine armchair nanotubes and a central scattering region made of $\sqrt{3}\times\sqrt{3}$ armchair carbon nanotubes. A gate voltage is applied in the central region. We consider typical $(8,8)$ armchair nanotubes at both parts, whose energy dispersions are respectively displayed in Figs.~\ref{fig08}(b) and \ref{fig08}(c). The intervalley coupling and valley-orbit coupling mechanism in $\sqrt{3}\times \sqrt{3}$ superlattices can manipulate valley polarization coherently in analogy to real spin for spintronics. Electrons propagating in the left pristine armchair carbon nanotube contain equivalent $K$ and $K'$ valley components. When the valley-unpolarized current in pristine armchair CNT enters the central $\sqrt{3} \times \sqrt{3}$ armchair nanotube region, these mechanisms break the balance between the two components by flipping electrons of K valley to K' valley or vice versa. Hence in the outgoing current to the right lead, one valley component is larger than the other. In another word, the current is valley-polarized. Similar to the spin polarization, a valley polarization function can be defined to evaluate the efficiency of the device. The valley polarization can be adjusted by external factors, such as bias voltage and gate voltage. The setup presented here can serve as a prototype valley field effect transistor, where valley-polarized current is turned on/off via applying the gate voltage as illustrated below.

The efficiency of this valley field effect transistor is characterized by the valley polarization function, which can be defined in terms of the valley-specified transmission function as:
\begin{equation}
P_{\rm V} = \frac{T_{\rm K} - T_{\rm K'}}{T_{\rm K} + T_{\rm K'}} \label{eq03}
\end{equation}
where $T_{\rm K}$ and $T_{\rm K'}$ are transmission functions of electrons belonging to $K$ and $K'$ valleys, respectively. To separate electrons from equivalent valleys, a simple and effective way is to consider their different group velocities. The group velocity, $v = \frac{1}{\hbar}\frac{\partial{E}}{\partial{k}}$, is related to the band structure, i.e., the slope of the band structures. When focusing on the first subband of $(8,8)$ pristine armchair carbon nanotube shown in Fig.~\ref{fig08}(b), it is clear that above the Fermi energy electron in $K$ valley has larger group velocity than that of $K'$ valley. The situation becomes reverse below the Fermi level. Thus, we can calculate the transmission function contributed from any specific valley under the Green's function frame. In the semi-infinite lead of pristine armchair carbon nanotube, the velocity of incident electrons is connected with the line width function $\Gamma_{\rm L}$ in the form of $\hbar {{\bf{v}}_{\rm L}}= {\bf{U}^\dag} \Gamma_{\rm L} {\bf{U}} = \widetilde{\Gamma}_{\rm L}$~\cite{velocity1,velocity2}. Here ${\bf{v}}_{\rm L}$ is a diagonal velocity matrix with nonzero diagonal elements contributed by electrons incoming from the left lead. $\bf U$ is a unitary transformation matrix ranked by eigenfunctions of $\Gamma_{\rm L}$, transforming it into a diagonal matrix $\widetilde{\Gamma}_{\rm  L}$. Obviously, there is an exact mapping between the incident electron velocities and the eigenvalues of line width function: ${{\bf{v}}_{\rm L}} = (1/ \hbar) \widetilde{\Gamma}_{\rm L}$. Considering only the propagating modes of the first subband of $(8,8)$ armchair carbon nanotubes [See Fig.~\ref{fig08}(b)], both ${{\bf{v}}_{\rm L}}$ and $\widetilde{\Gamma}_{\rm L}$ are $2 \times 2$ diagonal matrices. Incident electron from $K$ valley has larger velocity, corresponding to the larger one of the two eigenvalues of $\Gamma_{\rm L}$. Based on this analysis, we can construct effective line width function using only the propagating $K$ and $K'$ valley modes\cite{velocity2,eigengamma}
\begin{equation}
\overline{\Gamma}_{\rm L,K/K'} = \Lambda_{\rm K/K'} \hbar {\widetilde{\Gamma}_{\rm L,K/K'}} {\Lambda}^\dag_{\rm K/K'}  \nonumber,
\end{equation}
where $\Lambda_{\rm K,K'}$ is the eigenfunction of $\Gamma_{\rm L}$ for $K/K'$ valley mode. $\overline{\Gamma}_{\rm R,K/K'}$ of the right lead can be produced similarly. Following Eq.~(\ref{eq02}), the valley-specified transmission function can be straightforwardly expressed as
\begin{equation}
T_{\rm K/K'} = {\rm Tr}[\overline{\Gamma}_{\rm L,K/K'} G^r \overline{\Gamma}_{\rm R,K/K'} G^a] \nonumber
\end{equation}
Then one can calculate the valley polarization $P_{\rm V}$ via Eq.~(\ref{eq03}).

\begin{figure}[tbp]
\centering
\includegraphics[width=\columnwidth]{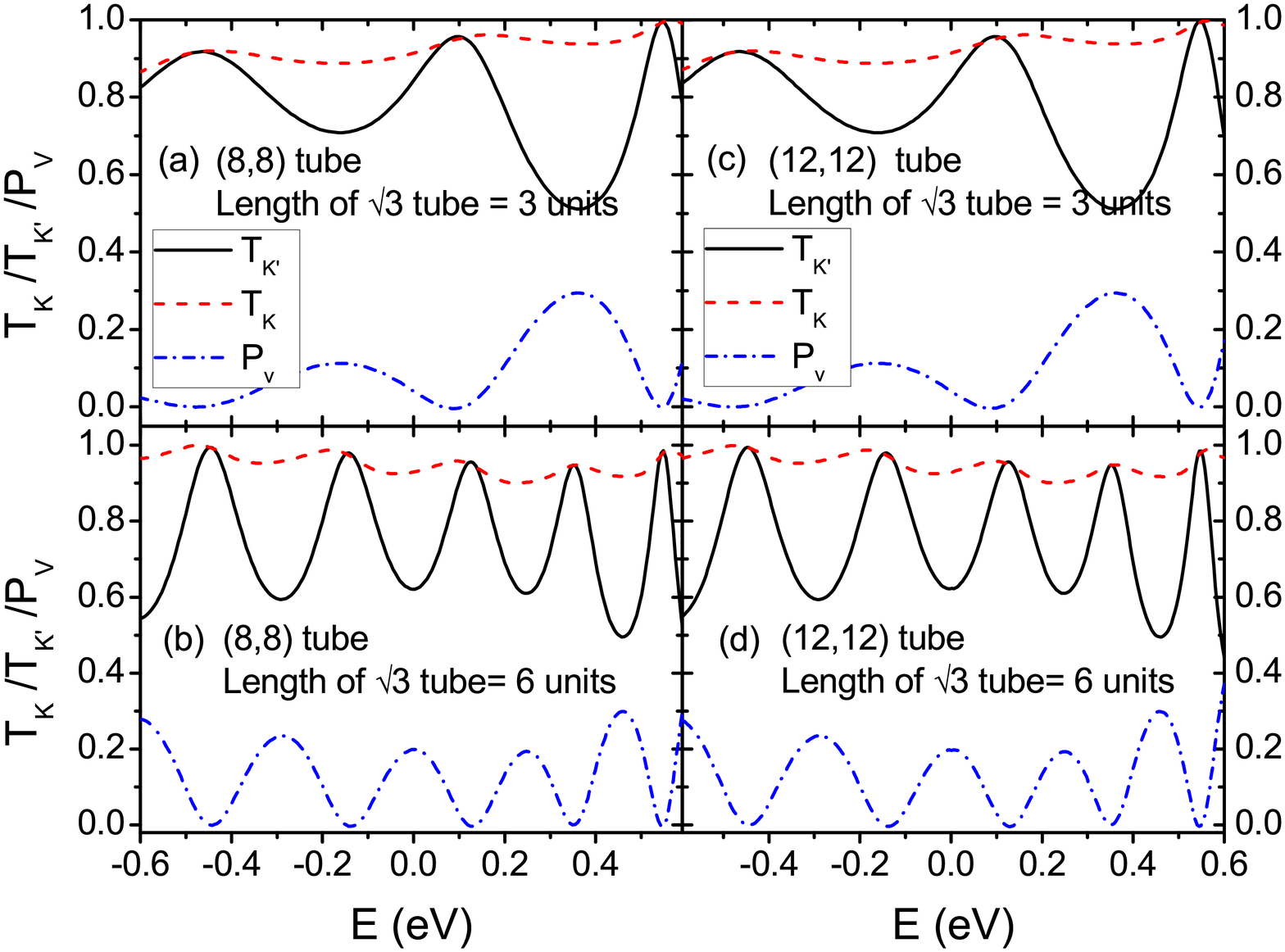}
\caption{Panels (a) and (b): Valley-specified transmission functions and valley polarization versus energy of $(8,8)$ armchair carbon nanotube valley-field-effect-transistor. Lengths of the $\sqrt{3}$ armchair carbon nanotube are respectively 3 and 6 units. Panels (c) and (d): The same functions for $(12,12)$ armchair carbon nanotube valley-field-effect-transistor. Gate voltage is set to be zero in the calculations. }
\label{fig09}
\end{figure}

We first evaluate the valley-specified transmission coefficient $T_{\rm K/\rm K'}$ and valley polarization efficiency $P_{\rm V}$ as function of the electron energy. The system under investigation is armchair carbon nanotube-based valley-field-effect-transistor as schematically plotted in Fig.~\ref{fig08}(a). The numerically calculated transport properties are exhibited in Fig.~\ref{fig09}. The energy interval of interest is the first subband of pristine armchair carbon nanotube. For the $(8,8)$ armchair system with 3 units length of $\sqrt{3}\times \sqrt{3}$ armchair carbon nanotube, both $T_{\rm K}$ and $T_{\rm K'}$ are continuous functions of the electron energy as shown in Fig.~\ref{fig09}(a). One can see that $T_{\rm K}$ is rather smooth and larger than 0.8 in the focused energy regime, while the magnitude of $T_{\rm K'}$ changes more abruptly and exhibits a fluctuating pattern. This observation reveals that incident electrons from $K$ valley of pristine nanotube are less affected by the central $\sqrt{3}\times\sqrt{3}$ armchair carbon nanotube. And this fact holds for all systems considered in Fig.~\ref{fig09}. The calculated valley polarization $P_{\rm V}$ is very small below the Fermi level and grows with the increasing electron energy. $P_{\rm V}$ also fluctuates like $T_{\rm K'}$ in the whole region, and its maximum reaches about 0.3 for our considered system.

When the length of central scattering region increases from 3 units to 6 units, $T_{\rm K}$ is still smooth as plotted in Fig.~\ref{fig09}(b), but $T_{\rm K'}$ fluctuates more frequently in the same energy window. Therefore, the resulting valley polarization vibrates with energy for both below and above the Fermi level. This result suggests that, by increasing the length of central region, one can realize a more effective manipulation of valley polarization in a relatively small energy range. Considering a larger system, such as $(12,12)$ armchair carbon nanotube valley-field-effect-transistor, an interesting and important observation is: both $T_K$ and $T_{K'}$, as well as $P_v$, are identically the same as those in $(8,8)$ system in the energy window of $[-0.6,0.6]$~eV, as shown in Fig.~\ref{fig09}(c) and \ref{fig09}(d). We have examined setups from $(6,6)$ to $(12,12)$ carbon nanotube systems and reached the following conclusion: as long as the electron energy is in the first subband of the pristine armchair carbon nanotube, both the valley-specified transmission functions and valley polarization are independent of the circumference of the system. This fantastic property guarantees a great freedom in fabricating such kind of valley-field-effect-transistor, since the device's performance is independent of its transverse dimension.

\begin{figure}[bp]
\centering
\includegraphics[width=\columnwidth]{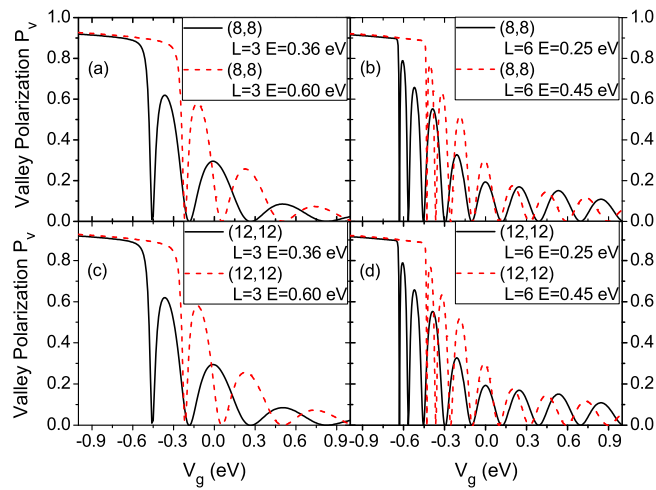}
\caption{Panels (a) and (b): Valley polarization versus gate voltage of $(8,8)$ armchair carbon nanotube valley-field-effect-transistor at several electron energies. 'L' in the legends stands for the lengths of the $\sqrt{3}$ armchair carbon nanotube, which are respectively 3 and 6 units in the calculation. Panels (c) and (d): The same functions for $(12,12)$ tube systems. }
\label{fig10}
\end{figure}

Secondly, we investigate the influence of the gate voltage on the device performance. The gate voltage is applied on the $\sqrt{3}\times \sqrt{3}$ armchair carbon nanotube region, which serves as the valley-processing unit. In our calculation, gate voltage simply shifts the on-site energies of the affected atoms, \textit{i.e.}, diagonal elements of their Hamiltonian. Electron energy is kept at the first subband of pristine armchair carbon nanotube. The valley polarization $P_{\rm V}$ as a function of gate voltage $V_{\rm g}$ is calculated at two energy points and different system sizes as shown in Fig.~\ref{fig10}. From Fig.~\ref{fig10}(a), one can find that the valley polarization fluctuatingly grows as the applied gate voltage decreases from positive to negative. For a $(8,8)$ valley-field-effect-transistor with 3-unit length of the $\sqrt{3} \times \sqrt{3}$ armchair carbon nanotube, the valley polarization can reach about 0.9 at the negative gate voltage, which indicates that the $K$ valley is almost fully polarized. The wide $P_{\rm V} > 0.9$ plateau shown in Fig.~\ref{fig10}(a) qualifies the device as a stable valley-polarized current generator that can operate in a broad gate voltage range.

More importantly, $P_{\rm V}$ jumps abruptly from below 0.02 to above 0.9 at $V_g \approx -0.5$~eV, showing a great on/off ratio of this valley-field-effect-transistor. The fluctuation of $P_{\rm V}$ exists at all electron energies. Comparing the results for $E=0.36 $~eV and $E=0.60$~eV, obviously high valley polarization can be easily achieved when the system works at larger electron energy, which only requires $V_g \approx -0.25$~eV to reach $P_{\rm V} > 0.9$ at $E=0.60$~eV. When increasing the length of the central regime to 6 units in the device, more fluctuations of $P_{\rm V}$ are revealed in Fig.~\ref{fig10}(b), suggesting a more effective gate modulation. But the high valley polarization plateau shrinks at all electron energies and a larger negative $V_{\rm g}$ is required for $P_{\rm V} > 0.9$. We also performed calculations on a $(12,12)$ armchair carbon nanotube system and the corresponding results are displayed in Fig.~\ref{fig10}(c) and \ref{fig10}(d). It is found again that the device performance is independent of the diameter of the nanotubes.

\begin{figure}[tbp]
\centering
\includegraphics[width=\columnwidth]{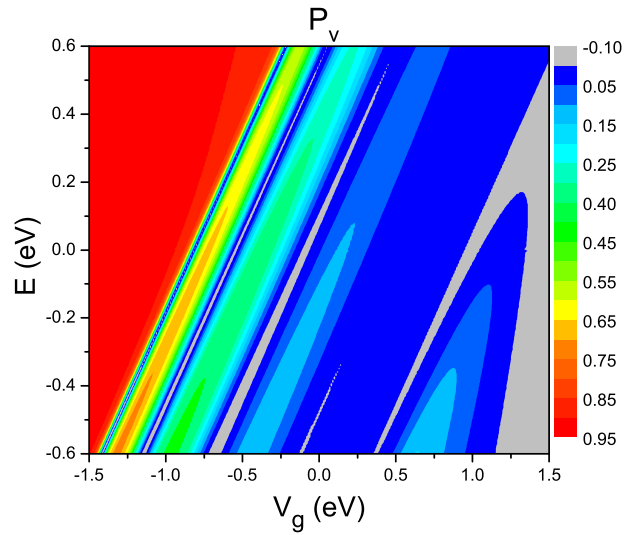}
\caption{Valley polarization function $P_{\rm V}$ of a $(8,8)$ armchair carbon nanotube valley-field-effect-transistor with 3 units length, in terms of the electron energy and gate voltage.}
\label{fig11}
\end{figure}

We summarize this part with the working map of a $(8,8)$ armchair carbon nanotube based valley-field-effect-transistor. Its valley polarization $P_{\rm V}$ as functions of electron energy and gate voltage are displayed in Fig.~\ref{fig11}. $P_{\rm V}$ function of this device can be effectively tuned by applying an external gate voltage. The best working zone of this device is the large red triangular area highlighted in Fig.~\ref{fig11}, where the $K$ valley is nearly fully polarized. The valley-field-effect-transistor has a stable output in this zone, and the on/off ratio is guaranteed by the narrow blue region adjacent to the red zone, whose valley polarization is below 0.05. The larger the electron energy, the easier to get fully valley-polarized current. The performance of this device is the same for systems with different circumferences, as long as the energy is within the first subband of pristine armchair carbon nanotube.

\section{Conclusion}\label{sec4}

In conclusion, we have numerically investigated the electronic properties of typical $\sqrt{3} \times \sqrt{3}$ graphene nanoribbon and nanotube structures. We show that all the $\sqrt{3} \times \sqrt{3}$ nanostructures are metallic materials. Both the zigzag and armchair ribbons have finite-size energy gap below the Fermi energy more than 1 eV. Double-degeneracy in energy levels instead of energy gaps are found in the spectra of both armchair and zigzag carbon nanotubes. In small $\sqrt{3}\times \sqrt{3}$ carbon nanotubes, there is a large energy range showing linear dispersion, which leads to the Klein tunneling-like behavior: electron transmission through the tube is quantized even in the presence of a potential barrier, regardless of the barrier length and height up to a few hundreds of meV. A valley-field-effect-transistor consisting of pristine and $\sqrt{3}\times\sqrt{3}$ armchair carbon nanotubes is proposed, which can be used to filter fully valley-polarized current and can be tuned by applying an external gate voltage or adjusting the length of the central scattering regime. As long as the electron energy is within the range of the first subband of pristine armchair carbon nanotubes, performance of this valley-field-effect-transistor is independent of the tube circumference.

\section{acknowledgments}
This work was financially supported by the NNSFC (Grants No. 11504240, No. 11574217, No. 11304205, and No. 11474265), NSF of SZU (Grant No. 201550). Y.R. and Z.Q. also acknowledge the financial supports from the China Government Youth 1000-Plan Talent Program, Fundamental Research Funds for the Central Universities (WK3510000001 and WK2030020027) and the National Key R \& D Program (Grant No. 2016YFA0301700) . The Supercomputing Center of USTC is gratefully acknowledged for the high-performance computing assistance.

\end{document}